\documentclass[11pt]{llncs}

by 2 \advance \topmargin -1.7in \leftmargin 210mm \advance
\leftmargin -\textwidth \divide \leftmargin by 2 \advance
\leftmargin

\usepackage{graphicx}

\begin{document}
\pagestyle{plain}

\title{On the Hopcroft's minimization algorithm}
\titlerunning{On the Hopcroft's minimization algorithm}
\author{Andrei P\u aun}
 \institute{ Department of Computer Science/IfM, Louisiana Tech University\\ P.O. Box 10348, Ruston, LA
 71272, USA\\
 Universidad Polit\'ecnica de Madrid - UPM,
Facultad de Inform\'atica\\ Campus de Montegancedo S/N, Boadilla del
Monte, 28660 Madrid, Spain \email{apaun@latech.edu}
 }
\maketitle

\begin{abstract}
We show that the absolute worst case time complexity for Hopcroft's
minimization algorithm applied to unary languages is reached only
for de Bruijn words. A previous paper by Berstel and Carton gave the
example of de Bruijn words as a language that requires O(n log n)
steps by carefully choosing the splitting sets and processing these
sets in a FIFO mode. We refine the previous result by showing that
the Berstel/Carton example is actually the absolute worst case time
complexity in the case of unary languages. We also show that a LIFO
implementation will not achieve the same worst time complexity for
the case of unary languages. Lastly, we show that the same result is
valid also for the cover automata and a modification of the
Hopcroft's algorithm, modification used in minimization of cover
automata.
\end{abstract}

\section{Introduction}

This work is a continuation of the result reported by Berstel and
Carton in \cite{berstel_ciaa02}. There they showed that Hopcroft's
algorithm
 requires O(n
log n) steps when considering the example of de Bruijn words (see
\cite{bruijn}) as input. The setting of the paper
\cite{berstel_ciaa02} is for languages over an unary alphabet,
considering the input languages having the number of states a power
of 2 and choosing ``in a specific way" which set to become a
splitting set in the case of ties. In this context, the previous
paper showed that one needs $O(n\ log\ n)$ steps for the algorithm
to complete, which is reaching the theoretical asymptotic worst case
time complexity for the algorithm as reported in
\cite{hopcroft,HopUll,gries,Knuutila} etc.

We were interested in investigating further this aspect of the
Hopcroft's algorithm, specifically considering the setting of unary
languages, but for a stack implementation in the algorithm. Our
effort has lead to the observation that when considering the worst
case for the number of steps of the algorithm (which in this case
translates to the largest number of states appearing in the
splitting sets), a LIFO implementation indeed outperforms a FIFO
strategy as suggested by experimental results on random automata as
reported in \cite{CIAA06}. One major observation/clarification that
is needed is the following: we do not consider the asymptotic
complexity of the run-time, but the actual number of steps. For the
current paper when comparing $n\ log\ n$ steps and $n\ log(n-1)$
steps we will say that $n\ log\ n$ is worse than $n\ log(n-1)$, even
though when considering them in the framework of the asymptotic
complexity (big-O) they have the same complexity, i.e. $n\ log\ n\in
\Theta(n\ log(n-1))$.

We give some definitions, notations and previous results in the next
section, then we give a brief description of the algorithm discussed
and its features in Section \ref{hop}, Section \ref{worst} describes
the properties for the automaton that reaches worst possible case in
terms of steps required for the algorithm (as a function of the
initial number of states of the automaton). We then briefly touch
upon the case of cover automata minimization with a modified version
of the Hopcroft's algorithm in Section \ref{cover} and conclude by
giving some final remarks in the Section \ref{fin}.

\section{Preliminaries}
\label{prelim}

We assume the reader is familiar with the basic notations of formal
languages and finite automata, see for example the excellent work by
Hopcroft, Salomaa or Yu \cite{HopUll,ASalaa,syu}. In the following
we will be denoting the cardinality of a finite set $T$ by $|T|$,
the set of words over a finite alphabet $\Sigma$ is denoted
$\Sigma^*$, and the
 empty word is $\lambda$.
The length of a word $w\in \Sigma^*$ is denoted with $|w|$. We
define $\Sigma^{l}=\{w \in \Sigma^* \mid |w|=l\}$, $\Sigma^{\leq
l}=\displaystyle\bigcup_{i=0}^l \Sigma^i$, and $\Sigma^{<
l}=\displaystyle\bigcup_{i=0}^{l-1} \Sigma^i$.

A deterministic finite automaton (DFA) is a quintuple
$A=(\Sigma,Q,\delta, q_0,F)$ where $\Sigma$ is a finite set of
symbols, $Q$ is a finite set of states, $\delta:Q \times \Sigma
\longrightarrow Q$ is the transition function, $q_0$ is the start
state, and $F$ is the set of final states. We can extend $\delta$
from $Q \times \Sigma $ to $Q \times  \Sigma^* $ by
$\overline{\delta}(s,\lambda)=s,$
$\overline{\delta}(s,aw)=\overline{\delta}(\delta(s,a),w).$ We
usually denote the extension $\overline{\delta}$ of $\delta$ by
$\delta$.

The language recognized by the automaton $A$ is $L(A)=\{w\in
\Sigma^*\mid \delta(q_0,w)\in F\}$. For simplicity, we assume that
$Q=\{0,1,\ldots,|Q-1|\}$ and $q_0=0$. In what follows we assume that
$\delta$ is a total function, i.e., the automaton is  complete.

For a DFA $A=(\Sigma,Q,\delta,q_0,F)$, we can always assume, without
loss of generality, that $Q=\{0,1,\ldots,n-1\}$ and $q_0=0$; we will
use this idea every time it is convenient for simplifying
 our notations.
 If $L$ is finite, $L=L(A)$ and $A$ is complete, there is at least
 one state, called the sink state or dead state, for which
 $\delta(sink,w)\notin F$, for any $w\in \Sigma^*$.
If $L$ is a finite language, we denote by $l$ the maximum among the
length of words in $L$.

\begin{definition}
A language $L'$ over $\Sigma$ is called a cover language for the
finite language
 $L$ if
$L'\cap\Sigma^{\leq l} = L$. A deterministic finite cover automaton
(DFCA) for $L$ is a  deterministic finite automaton (DFA) $A$, such
that the language accepted by $A$ is a cover language of $L$.
\end{definition}

\begin{definition}
Let $A=(Q, \Sigma , \delta , 0 , F)$ be a DFA and $L=L(A)$. We say
that $p\equiv_A q $ (state $p$ is equivalent to $q$ in $A$)
 if for every $w\in \Sigma^{*}$,
$\delta(s,w)\in F$ iff $\delta(q,w)\in F$.
\end{definition}

The right language of state $p\in Q$ for a DFCA
$A=(Q,\Sigma,\delta,q_0,F)$ is $R_p=\{w\mid \delta(p,w)\in F,
|w|\leq l-level_A(p)\}$.

\begin{definition}
Let $x,y\in \Sigma^*$. We define the following similarity relation
by: $x \sim_L y$ if for all $z\in \Sigma^*$ such that
$xz,yz\in\Sigma^{\leq l}$, $xz\in L$ iff $yz\in L$, and we write
$x\not\sim_L y$ if $x\sim_L y$ does not hold.
\end{definition}

\begin{definition}
Let $A = (Q,\Sigma,\delta,0,F)$ be a DFA (or a DFCA). We define, for
each state $q\in Q$, $level(q)=\min\{|w| \mid \delta(0,w)=q\}$.
\end{definition}

\begin{definition}
\label{def_states} Let $A=(Q, \Sigma , \delta , 0 , F)$ be a DFCA
for $L$. We consider two states $p,\ q\in Q$ and
$m=\max\{level(p),level(q)\}$. We say that $p$ is similar with $q$
in $A$, denoted by $p \sim_A q$, if for every $w\in \Sigma^{\leq
l-m}$, $\delta(p,w)\in F$ iff $\delta(q,w)\in F$. We say that two
states are dissimilar if they are not similar.
\end{definition}

If the automaton is understood, we may omit the subscript $A$.

\begin{lemma}
\label{eq_Lq=Ls} Let $A=(Q,\Sigma,\delta,0,F)$ be a DFCA of a finite
language $L$. Let $level(p)=i$, $level(q)=j$, and $m=\max\{i,j\}$.
If $p\sim_A q$, then $R_p\cap \Sigma^{\leq l-m}=R_q\cap \Sigma^{\leq
l-m}$.
\end{lemma}

\begin{definition}
A DFCA $A$ for a finite language is a minimal DFCA if and only if
any two
distinct states of $A$ are dissimilar. 
\end{definition}

Once two states have been detected as similar, one can merge the
higher level one into the smaller level one by redirecting
transitions. We refer the interested reader to \cite{wia98} for the
merging theorem and other properties of cover automata.

\section{Hopcroft's state minimization algorithm}\label{hop}

In \cite{hopcroft} it was described an elegant algorithm for state
minimization of DFAs. This algorithm was proven to be of the order
$O(n\ log\ n)$ in the worst case (asymptotic evaluation).

The algorithm uses a special data structure that makes the set
operations of the algorithm fast. We now give the description of the
algorithm as given for an arbitrary alphabet $A$ and working on an
automaton $(A,Q,\delta,q_0,F)$ and later we will restrict the case
to the unary languages.

\bigskip

 1: $P =\{F,\ Q-F\}$

 2: for all $a \in A$ do

 3: \hspace*{0.5cm} Add((min($F,\ Q-F), a), S$)

 4: while $S \not =\emptyset $ do

 5: \hspace*{0.5cm} get $(C, a)$ from $S$ \ \ (we extract $(C,a)$ according to the

 \hspace*{0.9cm} strategy associated with $S$: FIFO/LIFO/...)

 6: \hspace*{0.5cm} for each $B \in P$ split by $(C, a)$ do

 7: \hspace*{0.5cm}\hspace*{0.5cm}  $B'$, $B''$ are the sets resulting from splitting of $B$ w.r.t.  $(C,
 a)$

 8: \hspace*{0.5cm}\hspace*{0.5cm} Replace $B$ in $P$ with both $B'$ and $B''$

 9: \hspace*{0.5cm}\hspace*{0.5cm} for all $b \in A$ do

 10: \hspace*{0.5cm}\hspace*{0.5cm}\hspace*{0.5cm} if $(B, b) \in S$ then

 11: \hspace*{0.5cm}\hspace*{0.5cm}\hspace*{0.5cm}\hspace*{0.5cm}Replace $(B, b)$ by $(B', b)$ and $(B'', b)$ in $S$

 12: \hspace*{0.5cm}\hspace*{0.5cm}\hspace*{0.5cm}else

 13: \hspace*{0.5cm}\hspace*{0.5cm}\hspace*{0.5cm}\hspace*{0.5cm} Add((min$(B',B''), b), S)$
\bigskip

Where the splitting of a set $B$ by the pair $(C,a)$ (the line 6)
means that $\delta(B,a)\cap C\not=\emptyset$ and $\delta(B,a)\cap
(Q-C)\not=\emptyset$. Where by $\delta(B,a)$ we denote the set
$\{q\mid q=\delta(p,a),\ p\in B\}$. The $B'$ and $B''$ from line 7
are defined as the two subsets of $B$ that are defined as follows:
$B'=\{b\in B\mid \delta(b,a)\in C\}$ and $B''=B-B'$.
\medskip

It is useful to explain briefly the algorithm: we start with the
partition $P=\{F,Q-F\}$ and one of these two sets is then added to
the splitting sequence $S$. The algorithm proceeds in splitting
according to the current splitting set retrieved from $S$, and with
each splitting of a set in $P$ the splitting sets stored in $S$
grows (either through instruction 11 or instruction 13). When all
the splitting sets from $S$ are processed, and $S$ becomes empty,
then the partition $P$ shows the state equivalences in the input
automaton:  all the states contained in a same set $B$ in $P$ are
equivalent. Knowing all equivalences, one can easily minimize the
automaton by merging all the sets in the same set in the final
partition $P$.

We note that there are three levels of ``nondeterminism" in the
algorithm: the ``most visible one" is the strategy for processing
the list stored in $S$: as a queue, as a stack, etc. The second and
third levels of nondeterminism in the algorithm appear when a set
$B$ is split into $B'$ and $B''$. If $B$ is not present in $S$, then
the algorithm is choosing which set $B'$ or $B''$ to be added to
$S$, choice that is based on the minimal number of states in these
two sets. In the case when both $B'$ and $B''$ have the same number
of states, then we have the second ``nondeterministic" choice. The
third such choice appears when the splitted set $(B,a)$ is in the
list $S$; then the algorithm mentions the replacement of $(B,a)$ by
$(B',a)$ and $(B'',a)$ (line 11). This actually is implemented in
the following way: $(B'',a)$ is replacing $(B,a)$ and $(B',a)$ is
added to the list $S$ (or vice-versa). Since we saw that the
processing strategy of $S$ matters, then also the choice of which
$B'$ or $B''$ is added to $S$ and which one replaces the previous
location of $(B,a)$ matters in an actual implementation.

In the original paper \cite{hopcroft} and later in \cite{gries}, and
\cite{Knuutila} when describing the complexity of the algorithm, the
authors showed that the algorithm is influenced by the number of
states that appear in the sets processed by $S$. Intuitively, that
is why the smaller of the $B'$ and $B''$ is inserted in $S$ in line
13, and this makes the algorithm sub-quadratic. In the following we
will focus on exactly this issue of the number of states appearing
in sets processed by $S$.

\section{Worst case scenario for unary languages}\label{worst}

Let us start the discussion by making several observations and
preliminary clarifications: we are discussing about languages over
an unary alphabet. To make the proof easier, we restrict our
discussion to the automata having the number of states a power of 2.
The three  levels of nondeterminism are clarified in the following
way: we assume that the processing of $S$ is based on a FIFO
approach, we also assume that there is a strategy of choosing
between two just splitted sets having the same number of elements in
such a way that the one that is added to the queue $S$ makes the
third nondeterminism non-existent. In other words, no splitting of a
set already in $S$ will take place. We denote by $S_{w},\
w\in\{0,1\}^*$ the set of states $p\in Q$ such that
$\delta(p,a^{i-1})\in F$ iff $w_i=1$ for $i=1..|w|$, where
$\delta(p,a^0)$ denotes $p$. As an example, $S_1=F$, $S_{110}$
contains all the final states that are followed by a final state and
then by a non-final state and $S_{00000}$ denotes the states that
are non-final and are followed in the automaton by four more
non-final states.

Let us assume that such an automaton with $2^n$ states is given as
input for the minimization algorithm described in the previous
section. We note that since we have only one letter in the alphabet,
the states $(C,a)$ from the list $S$ can be written without any
problems as $C$, thus the list $S$ (for the particular case of unary
languages) becomes a list of sets of states. So let us assume that
the automaton $(\{a\},Q,\delta,q_0,F)$ is given as the input of the
algorithm, where $|Q|=2^n$. The algorithm proceeds by choosing the
first splitter set to be added to $S$. The first such set will be
chosen between $F$ and $Q-F$ based on their number of states. Since
we are interested in the worst case scenario for the algorithm, and
the algorithm run-time is influenced by the total number of states
that will appear in the list $S$ throughout the running of the
algorithm (as shown in \cite{hopcroft}, \cite{gries},
\cite{Knuutila} and mentioned in \cite{berstel_ciaa02}), it is clear
that we want to maximise the sizes (and their numbers) of the sets
that are added to $S$. It is time to give a Lemma that will be
useful in the following.

\begin{lemma}\label{number}
For deterministic automata over unary languages, if a set $R$ with
$|R|=m$ is the current splitter set, then $R$ cannot add to the list
$S$ sets containing more than $m$ states.
\end{lemma}

\begin{proof}
The statement of the lemma is saying that for all the sets $B_i$
from the current partition $P$ such that $\delta(B_i,a)\cap
R\not=\emptyset$ and $\delta(B_i,a)\cap (Q-R)\not=\emptyset$. Then
$\sum_{i} |B'_i|\le m$, where $B'_i$ is the smaller of the two sets
that result from the splitting of $B_i$ with respect to $R$.

We have only one letter in the alphabet, thus the number of states
$q$ such that $\delta(q,a)\in R$ is at most $m$. Each $B_i'$ is
chosen as the set with the smaller number of states when splitting
$B_i$ thus $|B'_i|\le |\delta(B_i,a)\cap R|$ which implies that
$\sum_{i} |B'_i|\le \sum_{i}|\delta(B_i,a)\cap R|=|(\bigcup_{i}
\delta(B_i,a))\cap R|\le |R|$  (because all $B_i$ are disjoint).

Thus we proved that if we start splitting according to a set $R$,
then the new sets added to $S$ contain at most $|R|$ states. \qed
\end{proof}

Coming back to our previous setting, we have the automaton given as
input to the algorithm and we have to find the smaller set between
$F$ and $Q-F$. In the worst case (according to Lemma \ref{number})
we have that $|F|=|Q-F|$, as otherwise, fewer than $2^{n-1}$ states
are contained in the set added to $S$ and thus less states will be
contained in the sets added to $S$ in the second stage of the
algorithm, and so on.

At this step either $F=S_1$ or $Q-F=S_0$ can be added to $S$ as they
have the same number of states.  Either one that is added to the
queue $S$ will split the partition $P$ in the worst case scenario in
the following four possible sets $S_{00},S_{01},S_{10},S_{11}$, each
with $2^{n-2}$ states. This is true as by splitting the sets $F$ and
$Q-F$ in sets with sizes other than $2^{n-2}$, then according with
Lemma \ref{number} we will not reach the worst possible number of
states in the queue $S$ and also splitting only $F$ or only $Q-F$
will add to $S$ only one set of $2^{n-2}$ states not two of them.

All this means that half of the non-final states go to a final state
($|S_{01}|=2^{n-2}$) and the other half go to a non final state
($S_{00}$). Similarly, for the final states we have that $2^{n-2}$
of them go to a final state ($S_{11}$) and the other half go to a
non-final state. The current partition at this step 1 of the
algorithm is $P=\{S_{00},S_{01},S_{10},S_{11}\}$ and the splitting
sets are one of the $S_{00}, S_{01}$ and one of the $S_{10},
S_{11}$. Let us assume that it is possible to chose the splitting
sets to be added to the queue $S$ in such a way so that no splitting
of another set in $S$ will happen, (chose in this case for example
$S_{10}$ and $S_{00}$). We want to avoid splitting of other sets in
$S$ since if that happens, then smaller sets will be added to the
queue $S$ by the splitted set in $S$
(see such a choice of splitters described in
\cite{berstel_ciaa02}).

We have arrived at step 2 of the processing of the algorithm, since
these two sets from $S$ are now processed, in the worst case they
will be able to add to the queue $S$ at most $2^{n-2}$ state each by
splitting each of them two of the four current sets in the partition
$P$. Of course, to reach this worst case, we need them to split
different sets, thus in total we obtain eight sets in the partition
$P$ corresponding to all the possibilities:
$P=\{S_{000},S_{001},S_{010},S_{011},S_{100},S_{101},S_{110},S_{111}\}$
having $2^{n-3}$ states each. Thus four of these sets will be added
to the queue $S$. And we could continue our reasoning up until the
$i$-th step of the algorithm:

We now have $2^{i-1}$ sets in the queue $S$, each having $2^{n-i}$
states, and the partition $P$ contains $2^i$ sets $S_w$
corresponding to all the words $w$ of the length $i$. Each of the
sets in the splitting queue is of the form $S_{x_1x_2\dots x_i}$,
then a set $S_{x_1x_2x_3\dots x_i}$ can only split at most two other
sets $S_{x_2x_3\dots x_{i-1}0}$ and $S_{x_2x_3\dots x_{i-1}1}$ from
the partition $P$. In the worst case all the level $i$ sets in the
splitting queue are not splitting a set already in the queue, and
split 2 distinct sets in the partition $P$, making the partition at
step $i+1$ the set $P=\{S_w\mid |w|=i+1\}$, and each such $S_w$
having exactly $2^{n-i-1}$ states. And in this way the process
continues until we arrive at the $n$-th step. If the process would
terminate before the step $n$, of course we would not reach the
worst possible number of states passing through $S$.

Let us now see the properties of an automaton that would obey such a
processing through the Hopcroft's algorithm. We started with $2^n$
states, out of which we have $2^{n-1}$ final and also $2^{n-1}$
non-final, out of the final states, we have $2^{n-2}$ that preceed
another final state ($S_{11}$), and also $2^{n-2}$ non-final states
that preceed other non-final states for $S_{00}$, etc. The strongest
restrictions are found in the final partition sets $S_w$, with
$|w|=n$ each have exactly one element, which means that all the
words of length $n$ over the binary alphabet can be found in this
automaton by following the transitions between states and having 1
for a final state and 0 for a non-final state. It is clear that the
automaton needs to be circular and following the pattern of de
Bruijn words. Such an automaton for $n=3$ was depicted in
\cite{berstel_ciaa02} as in the following Figure \ref{fig1}.

\begin{figure}[ht]
\begin{center}
\includegraphics[scale=0.5]{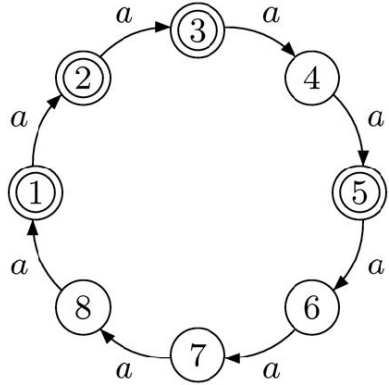}
\end{center}
\caption{A cyclic automaton of size 8 for the de Bruijn word
11101000.} \label{fig1}
\end{figure}

It is easy to see now that a stack implementation for the list $S$
will not be able to reach the maximum as smaller sets will be
processed before processing larger sets, which will lead to
splitting of sets already in the list $S$. Once this happens for a
set with $2^i$ states, then the number of states that will appear in
$S$ is decreased by at least $2^i$ because the splitted sets will
not be able to add as many states as a FIFO implementation was able
to do. We conjecture that in such a setting the LIFO strategy could
prove to make the algorithm liniar with respect to the size of the
input, if the aforementioned third level of nondeterminism is set to
add the smaller set of $B',\ B''$ to the stack and $B$ to be
replaced by the larger one. We proved the following result:

\begin{theorem}
The absolute worst case run-time complexity for the Hopcroft's
minimization algorithm for unary languages is reached when the
splitter list $S$ in the algorithm is following a FIFO strategy and
only for automata following de Bruijn words for size $n$. In that
setting the algorithm will pass through the queue $S$ exactly $n
2^{n-1}$ states.
\end{theorem}

\section{Cover automata}\label{cover}

In this section we discuss briefly (due to the page restrictions
imposed on the size of the paper) about an extension to Hopcroft's
algorithm to cover automata. K\"orner reported at CIAA'02 a
modification of the Hopcroft's algorithm so that the resulting sets
in the partition $P$ will give the similarities between states with
respect to the input finite language $L$.

To achieve this, the algorithm is modified as follows: each state
will have its level computed at the start of the algorithm; each
element added to the list $S$ will have three components: the set of
states, the alphabet letter and the current length considered. We
start with $(F,a,0)$ for example. Also the splitting of a set $B$ by
$(C,a,l_1)$ is defined as before with the extra condition that we
ignore during the splitting the states that have their level+$l_1$
greater than $l$ ($l$ being the longest word in the finite language
$L$). Formally we can define the sets $X=\{p\mid \delta(p,a)\in C, \
level(p)+l_1< l\}$ and $Y=\{p\mid \delta(p,a)\not\in C, \
level(p)+l_1< l\}$. Then a set $B$ will be split only if $B\cap
X\not=\emptyset$ and $B\cap Y\not=\emptyset$.

The actual splitting of $B$ ignores the states that have levels
higher than or equal with $l-l_1$. This also adds a degree of
nondeterminism to the algorithm when such states appear. The
algorithm proceeds as before to add the smaller of the newly
splitted sets to the list $S$ together with the value $l_1+1$.

Let us now consider the same problem as in \cite{berstel_ciaa02},
but in this case for the case of DFCA minimization through the
algorithm described in \cite{neamtu}. We will consider the same
example as before, the automata based on de Bruijn words as the
input to the algorithm (we note that the modified algorithm can
start directly with a DFCA for a specific language, thus we can have
as input even cyclic automata). We need to specify the actual length
of the finite language that is considered and also the starting
state of the de Bruijn automaton (since the algorithm needs to
compute the levels of the states). We can choose the length of the
longest word in $L$ as $l=2^n$ and the start state as $S_{111...1}$.
For example, the automaton in figure \ref{fig1} would be a cover
automaton for the language $L=\{0,1,2,4,8\}$ with $l=8$ and the
start state $q_0=1$. Following the same reasoning as in
\cite{berstel_ciaa02} but for the case of the new algorithm with
respect to the modifications, we can show that also for the case of
DFCA a queue implementation (as specifically given in \cite{neamtu})
seems a choice worse than a LIFO strategy for $S$. We note that the
discussion is not a straight-forward extension of the work reported
by Berstel in \cite{berstel_ciaa02} as the new dimension added to
the sets in $S$, the length and also the levels of states need to be
discussed in detail. We will give the details of the construction
and the step-by-step discussion of this fact in the journal version
of the paper.

\section{Final Remarks}
\label{fin}

We showed that at least in the case of unary languages, a stack
implementation is more desirable than a queue for keeping track of
the splitting sets in the Hopcroft's algorithm. This is the first
instance when it was shown that the stack is out-performing the
queue. It remains open whether there are examples of languages (over
an alphabet containing at least two letters) which for a LIFO
approach would perform worse or as worse as the FIFO. Our conjecture
is that the LIFO implementation will always outperform a FIFO
implementation, which was also suggested by the experiments reported
in \cite{CIAA06}. As future work planned, it is worth mentioning our
conjecture that there is a strategy for processing a LIFO list $S$
such that the minimization of all the unary languages will be
realized in linear time by the algorithm. We also plan to extend the
current results to the case of the cover automata, although, the
discussion in that case proves to be more complicated by the levels
of the states and the forth nondeterminism that this introduces.

\end{document}